\documentclass[referee]{aa}

\usepackage{amsmath}
\usepackage{amssymb}
\usepackage{latexsym}
\usepackage{subfig,graphicx}
\usepackage{txfonts}
\usepackage{natbib}
\bibpunct{(}{)}{,}{}{,}{,}
\usepackage{color}
\usepackage[mathcal]{euscript} 

\begin{document}
\title{Chromospheric jets around the edges of sunspots}

\author{R. J. Morton}

\institute{Solar Physics and Space Plasma Research Centre
(SP$^2$RC), University of Sheffield, Hicks Building, Hounsfield
Road, Sheffield S3 7RH, UK,\\
\\email:[r.j.morton]@sheffield.ac.uk}

\date{Received /Accepted}
\abstract{}{Evidence is beginning to be put forward that
demonstrates the role of the chromosphere in supplying energy and
mass to the corona. We aim to asses the role of chromospheric jets
in active region dynamics. }{Using a combination of the {Hinode/SOT}
Ca II H and TRACE $1550$~{\AA} and $1600$~{\AA} filters we examine
chromospheric jets situated at the edge of a sunspot.}{Analysis
reveals a near continuous series of jets, that raise chromospheric
material into the low corona above a sunspot. The jets have average
rise speeds of $30$~km\,s$^{-1}$ and a range of
$10-100$~km\,s$^{-1}$. Enhanced emission observed at the jets
leading edge suggests the formation of a shock front. Increased
emission in TRACE bandpasses above the sunspot and the disappearance
of the jets from the Ca II filter suggests that some of the
chromospheric jet material is at least heated to $\sim0.1$~MK. The
evidence suggests that the jets could be a mechanism which provides
a steady, low-level heating for active region features.}{}

\titlerunning{Chromospheric jets}
\authorrunning{Morton}

\maketitle

\section{Introduction}
Dynamic jets of plasma observed in the solar chromosphere are
currently the subject of much interest. \cite{DEPetal2011} has
demonstrated the association between spicules and material heated to
EUV temperatures ($T\sim0.2-1.0$~MK), suggesting they play an
important role in supplying heated plasma to maintain the quiet
corona. \cite{MADetal2011} {provides further evidence for a
connection between coronal hole spicules and transition region
emission but find no corresponding EUV ($T>0.3$~MK) emission.} {The
importance of plasma jets in atmospheric dynamics is also
highlighted in observations of large solar jets}, which show how
multiple, small jets can launch large quantities of cool
(chromospheric) and heated plasma into the corona (e.g.,
\citealp{MOOetal2011}; \citealp{SCUetal2011};
\citealp{MORetal2012b}).

Further to this, recent observations have shown that these dynamic
jets support ubiquitous transverse waves, both at the limb in
spicules (\citealp{DEPetal2007}; \citealp{ZAQERD2009}), on-disk in
mottles (\citealp{KURetal2012}), plumes (\citealp{TIAetal2011b}) and
in larger jet structures (\citealp{CIRetal2007};
\citealp{LIUetal2009}; \citealp{VAGetal2009};
\citealp{MORetal2012}). The transverse waves in spicules appear to
have sufficient energy to power the solar wind and possibly heat the
quiet solar corona.

The behaviour and appearance of chromospheric jets is well
documented for the quiet Sun, the plage of active regions and
coronal holes. \cite{DEPetal2007a} give evidence for the existence
of two types of spicule, namely Type I and Type II. Type I are
quantified as jets which rise with speeds of $10-40$~km\,s$^{-1}$
and exhibit parabolic profiles in time-distance plots, indicating a
portion of the jet's mass that is ejected returns to the surface. On
the other hand, type II's are shown to accelerate while rising,
reaching speeds of $50-100$~kms$^{-1}$ and fade from the filter
without being seen to return to the surface. It is these events, so
far, that have been observed to have a coronal counterpart
(\citealp{DEPetal2011}). The presence of a fast moving jets has also
been reported on disk (\citealp{ROUetal2009}).

Although it is well known {that} spicule-like features occur in
active regions (e.g., plage regions - dynamic fibrils), details of
their behaviour at the foot points of coronal structures are scarce.
Due to the substantial differences between active region and quiet
Sun magnetic structures, it might be expected that there is a
difference between the jet phenomena that inhabit the two regions.
This idea is supported by observed differences between quiet
Sun/coronal hole spicules (\citealp{DEPetal2007a}) and dynamic
fibrils in plage regions (\citealp{DEPetal2003};
\citealp{TZIetal2004}; \citealp{DEPetal2007c}). We note that
small-scale jets (widths $\sim400$~km) have been identified in the
chromospheric penumbra of sunspots (\citealp{KATetal2007}), however,
these jets are formed along low-lying magnetic structures that
probably do not penetrate into the corona.

\begin{figure*}[!hbt]
\centering
\includegraphics[scale=0.7] 
{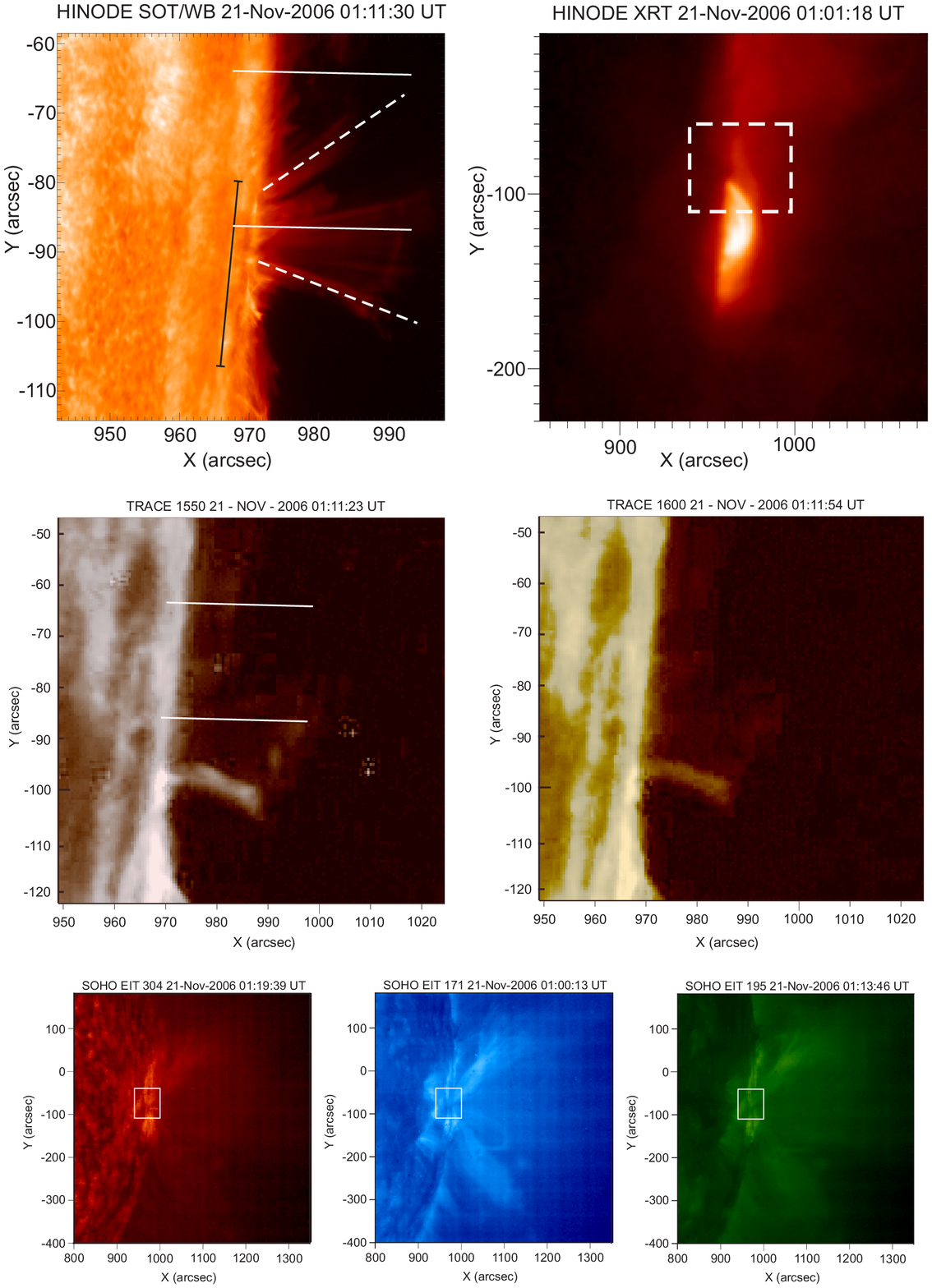}\caption{The field of view as observed by Hinode,
TRACE and SOHO EIT. The \textit{top left} hand image shows the
complete field of view for SOT. {The black line shows the position
of the sunspot umbra. Magnetic field lines, almost perpendicular to
the surface, emanate from the sunspot boundary and are outlined by
the Ca II emission}. The two dashed lines mark the positions of the
cross-cuts used in Fig.~\ref{fig:xt_plots}. The two solid lines mark
the position of cross-cuts used in Fig.~\ref{fig:emis}. The
\textit{top {right}} image shows a sub region of the XRT field of
view. The white box marks the SOT field of view. The \textit{middle
row} shows images of the sunspot at the limb as seen by TRACE
$1550$~{\AA} (\textit{left}) and 1600~{\AA} (\textit{right}). The
two solid lines in the $1550$~{\AA} image mark the position of
cross-cuts used in Fig.~\ref{fig:emis}. The \textit{bottom row}
shows the corresponding SOHO/EIT 304~{\AA}, 171~{\AA} and 195~{\AA}
images. The solid white boxes in each image mark the SOT field of
view.}\label{fig:fov}
\end{figure*}

{Closely related to the subject of jets are up-flows and down flows.
Flows have been observed above active regions in spectral lines
corresponding to transition region (TR) temperatures.
\cite{KJEetal1988} reported the presence of $5-20$~km\,s$^{-1}$
up-flows and $40-80$~km\,s$^{-1}$ down flows in C IV (1548~\AA),
which has a formation temperature of $\sim0.1$~MK. More recently,
\cite{TERIetal2008} used the SUMER (Solar Ultraviolet Measurements
of Emitted Radiation) instrument to investigate the TR above
sunspots, finding pervasive down flows on the order of
$10-30$~km\,s$^{-1}$ in Si IV (1403~\AA, $T\sim0.073$~MK). A number
of other studies have also shown that down flows are associated with
higher TR temperatures ($T\sim0.6-1.0$~MK) and high speed up flows
are seen at coronal temperatures ($T>1$~MK), e.g. \cite{MARetal2004,
MARetal2008}, \cite{BRYetal2010}, \cite{UGAWAR2011},
\cite{WARetal2011}, \cite{KAMetal2011}.}

{On the nature of the TR above sunspots, study of Hydrogen Lyman
line profiles with SUMER by \cite{TIAetal2009} revealed an
interesting insight. The authors suggested that the TR above a
sunspot is higher and more extended than the TR in plage regions.
They also conjectured there might be less chromospheric plasma above
sunspots.}

We aim to examine further the role that chromospheric phenomenon
play in the dynamics of active region structures connected with
sunspots. We investigate the properties of chromospheric jets that
surround a sunspot seen at the limb using Hinode and the Transition
Region And Coronal Explorer (TRACE). {Further, we demonstrate that
the TR is {higher} and more extended above the sunspot than in the
plage, as suggested by \cite{TIAetal2009}.} The chromosphere is also
extended and higher above the sunspot, suggesting it is less dense
in comparison to the plage chromosphere. Although it is challenging
to study the chromosphere/transition region above the sunspots,
imaging observations such as this one here provide an insight into
the role of the chromosphere in maintaining overlying active region
features.

\section{Data Analysis}
The data were obtained by Hinode Solar Optical Telescope (SOT -
\citealp{SUEetal2008}) at {01:11:00}~UT 21 November 2006 till
{02:00:00}~UT on the same day. Hinode/SOT viewed a region on the
west limb with the Ca II H broadband filter, which has a pixel size
of $0.054''$ but is diffraction limited to $0.2''$ ($145$~km). The
cadence of the data is $\sim8$~s. We performed the usual processing
routine for SOT data sets with \textit{fg\_prep.pro}. However, the
data still possess a significant drift over time. We correct for
this by tracking the limb over the time series and removing the
trend. All images (except those shown in Fig.~\ref{fig:fov}) are
then rotated so that the limb faces north.

In the time series, Hinode {was} following the active region 10923.
Hinode tracked its progress across the disk and the active region
has been the subject of many investigations (e.g.,
\citealp{KATetal2007}). In the data series used here, the sunspot is
about to cross the limb (Fig.~\ref{fig:fov}) allowing a novel
side-on view of the sunspots structure. The images clearly show
vertical and inclined magnetic fields emanating from the region into
the lower corona. A snapshot from XRT (\citealp{KANetal2008}) shows
that hot coronal plasma exists above the sunspot. {We do not use the
XRT data for analysis because the telescope was operating with a low
cadence ($3.5$~mins) at the time of interest. This cadence is much
too low to observe dynamic behaviour on the timescales observed in
Ca II H.}

\begin{figure*}[!tbp]
\centering
\includegraphics[scale=0.8] 
{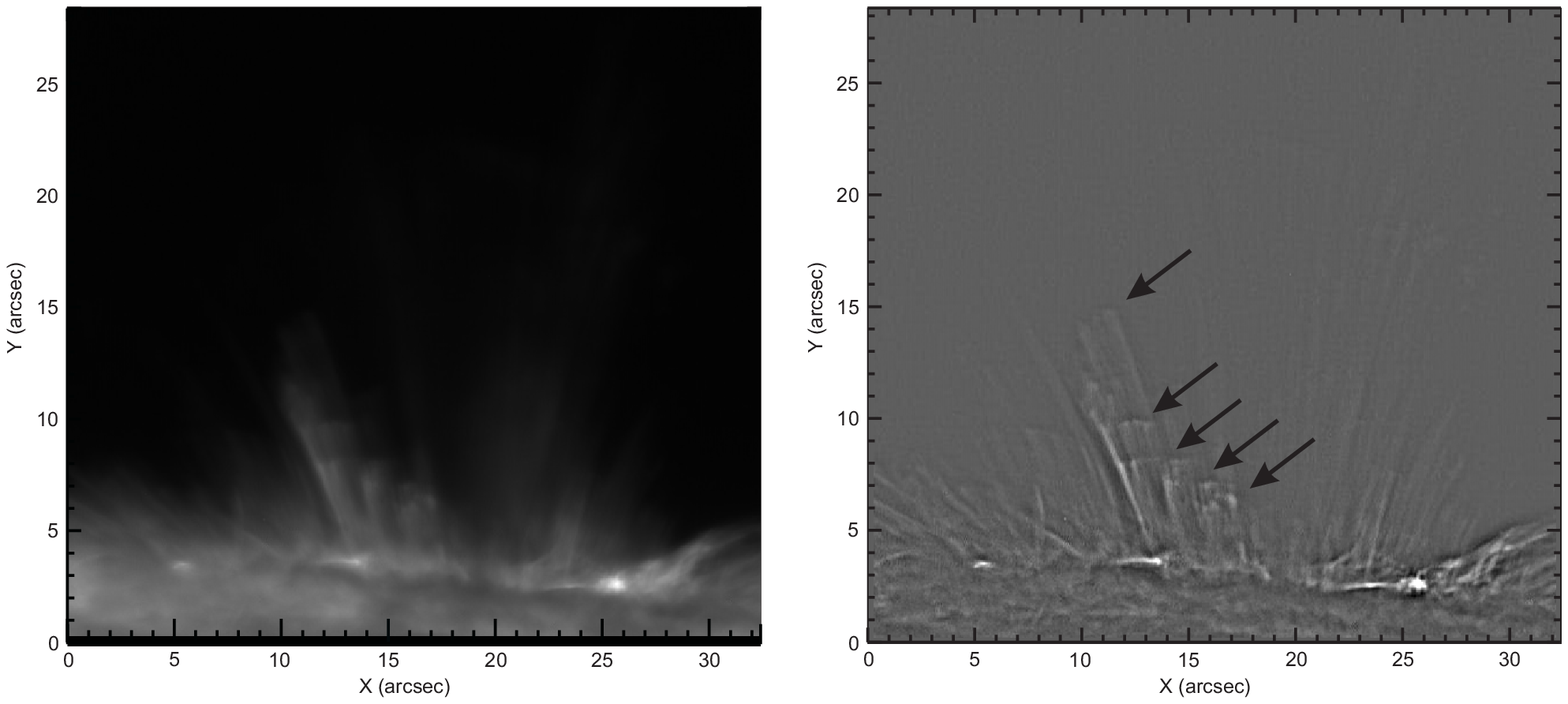}\caption{A Ca II H  image (\textit{left}) and an
unsharp masked version of the image. The magnetic field lines
surrounding the sunspot are highlighted by up-flowing plasma jets.
The jets are launched in groups along a number of field lines. These
discrete groups are highlighted by the arrows. The temporal
evolution can be seen in the movie available in the online
addition.}\label{fig:pack}
\end{figure*}

To emphasise magnetic structures and moving features we employ a
number of standard analysis techniques. The first is unsharp
masking, where, for each time frame, we subtract an image smoothed
with a boxcar function of width 10 by 10 pixels from the original
image. The result is seen in, e.g., Fig.~\ref{fig:pack}. Second, we
create running difference images of the original data by subtracting
the preceding time-frame from each image.

We also use data from the $1550$~{\AA} and $1600$~{\AA} filters on
board TRACE (\citealp{HANetal1999}) taken on the same day. The data
is subject to typical TRACE preparation routines. The data for these
bandpasses is sporadic in time and only covers certain parts of the
Hinode time series. Both these filters have significant
contributions from the continuum and from C IV
(\citealp{HANetal1998}) which has a formation temperature of
$T\sim0.1$~MK. We do not use the technique developed in
\cite{HANetal1998} for separating the continuum emission from C IV
emission as this is only suitable for intense solar features. {For
analysis between TRACE and Hinode/SOT images, we rebin the SOT data
so that pixels correspond to TRACE resolution and we align the
images via cross-correlation. }

{Further, data from the Solar and Heliospheric Observatory (SOHO)
Extreme ultraviolet Imaging Telescope (EIT - \citealp{DELetal1995})
is also used here and images are subject to the standard SOHO/EIT
preparation routines. The resolution of EIT is 2.56''
($\sim1000$~km), which is {much too} coarse to study the fine-scale
dynamics, so the EIT images will only provide context for the
Hinode/SOT and TRACE images.}

\section{Observational features}
On viewing the unsharp masked movie of the data, it is seen that the
plasma at the edge of the sunspot displays highly dynamic behaviour
(movie~1). Jets appear almost continuously from the chromosphere and
rise into the corona up to heights of $10-15$~Mm ($10-20$~arcsec).
The jets travel along fine-scale structures with widths on the order
of the diffraction limited resolution, i.e. $\sim150-200$~km. The
widths given are the values of the full-width half-maximums obtained
from a Gaussian fit to the structures cross-sectional intensity
profile.

Throughout the time series the plasma jets outline the magnetic
field revealing that it branches out almost symmetrically above the
sunspot. However, the jet features do not display the same behaviour
across the sunspot. It can be observed that the left and right hand
sides of the sunspot display much greater activity than at the
center. {Images from SOHO/EIT (Fig.~\ref{fig:fov}) reveal a {\lq
butterfly\rq} shaped active region emanating from the sunspot,
consisting of two sets of coronal loops highly inclined to the
vertical. The brightest emission at coronal heights in all EIT
channels is associated with the two sets of loops. The emission over
the center of the sunspot is fainter. The overall picture available
from Hinode and SOHO would suggest that the greater jet activity is
associated with the coronal loops}.
\begin{figure*}[!htbp]
\centering
\includegraphics[scale=0.8] 
{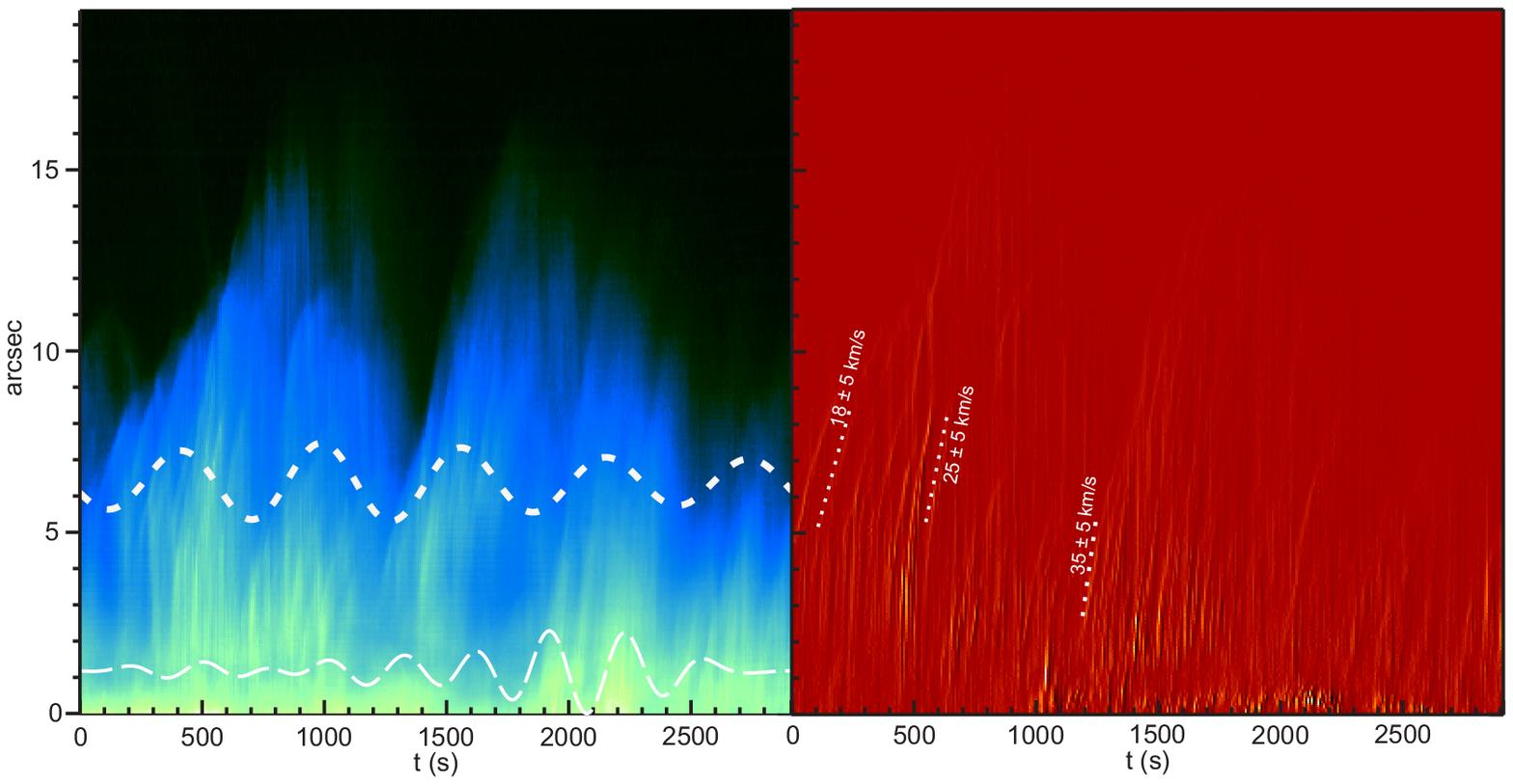} \vspace{0.5cm}
\includegraphics[scale=0.8] 
{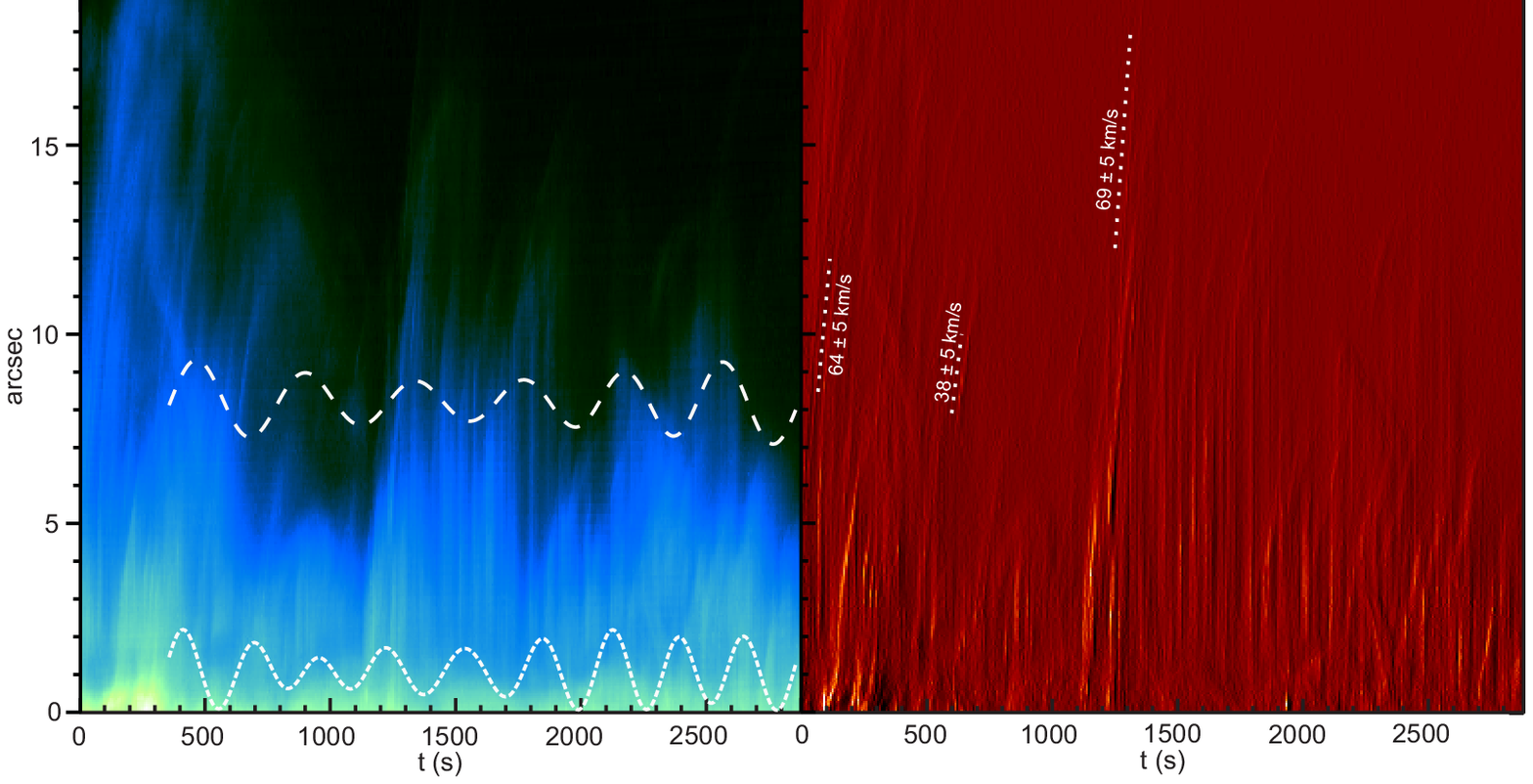} \caption{The \textit{left} hand column shows
time-distance diagrams take at the left (upper) and right (lower)
sides of the sunspot. The wavelet scales displaying the greatest
power are over plotted, with the zero crossings marking the height
at which the wavelet was taken. The \textit{right} hand columns
displays the time-distance diagrams taken from the running
difference movies. The dynamic up-flows are clearly seen. The
position of the cross-cuts are show in Fig.~\ref{fig:fov} and the
same cross-cut positions are used for the left and right columns.
}\label{fig:xt_plots}
\end{figure*}

The left-hand side of the sunspot shows near continuous jets, with
plasma emission reaching up to 15~Mm. The continuous nature of the
up-flow suggests a large amount of plasma is ejected into the
atmosphere but evidence for it returning to the surface is far less
obvious. Occasionally, receding jets can be observed but they appear
faint in intensity compared to the rising material. Continuous
up-flows are also seen on the right hand side of the sunspot.
However, intermittent, {large plasma ejections dominate the
behaviour of the jets, sending them higher into the atmosphere with
greater speeds} (e.g., Fig.~\ref{fig:xt_plots}).

The jets can also be observed to be generated in groups spanning
$\sim2000$~km and consist of a number of the resolution size
structures (Fig.~\ref{fig:pack}). These groups of jets make up the
majority of the observed up-flow on the left hand side of the
sunspot. To the best of our knowledge, no such collective behaviour
has been reported before in spicules/fibrils/mottles either in the
quiet or active Sun.

Next, we determine the rise speeds of the observed jets. We place a
cross-cut aligned with the jets upward trajectory (see
Fig.~\ref{fig:fov}) and produce time-distance diagrams from the
running difference movies of the data. Fig.~\ref{fig:xt_plots} shows
typical time-distance plots where we have marked on certain jet's
velocities. On calculating the velocities we have ignored projection
effects as we are unable to obtain any measurement of the
inclination along the line of sight. Note, a number of vertical
lines appear in the time-distance plots. These appear to be the
result of jets moving transversally across the sunspot, hence
passing through the cross-cut. {We do not identify any line in the
time-distance plots as a jet that shows emission over a length of
1~Mm in one time frame. This means we may also rule out some short
lived, fast-moving ($>125$~km\,s$^{-1}$) jets. Such jets should be
visible on viewing the unsharp-masked images, although we can
identify none.}

\begin{figure}[!tbp]
\centering
\includegraphics[scale=0.7] 
{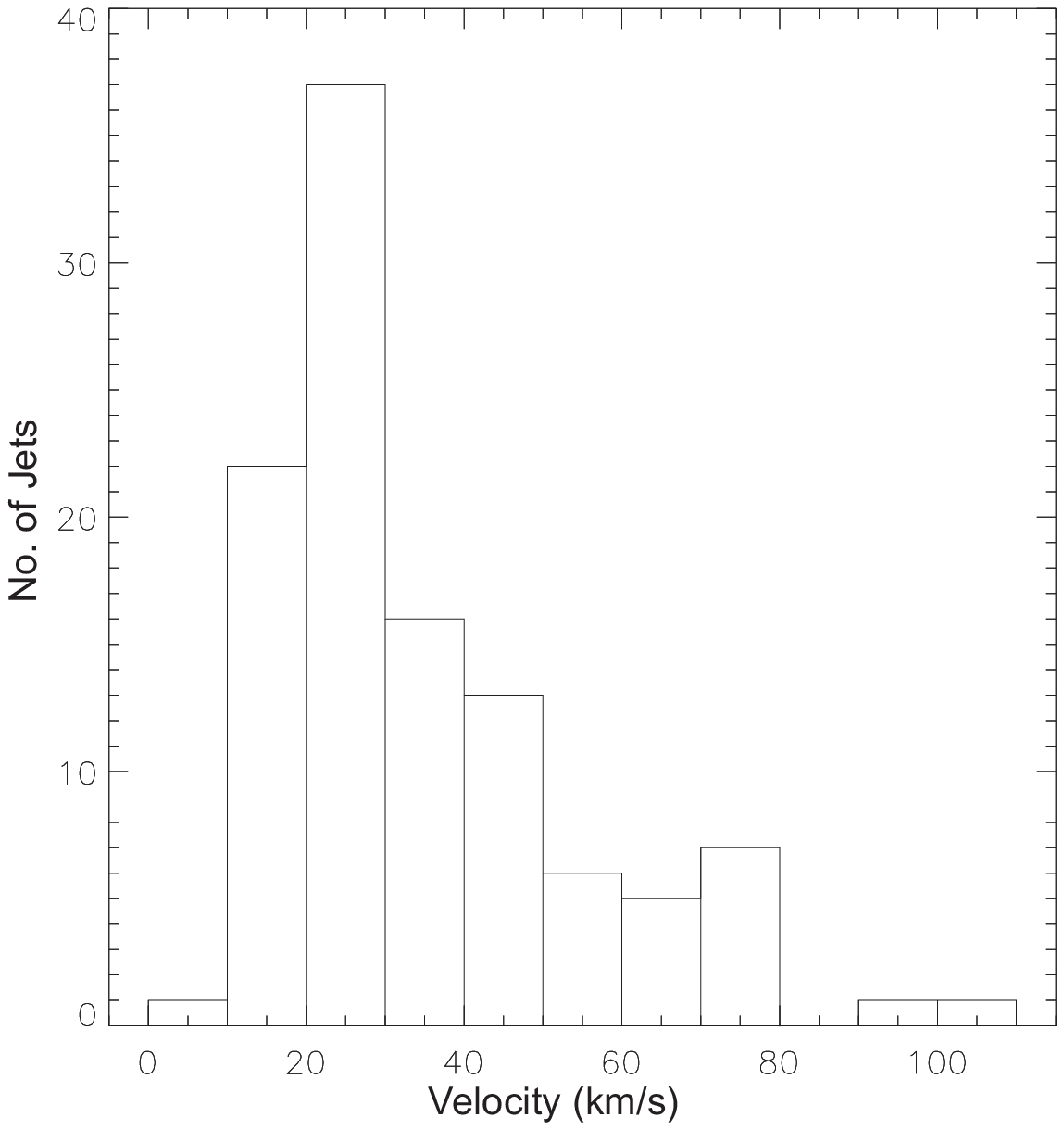}\caption{Histogram of jet velocities calculated
from time-distance diagrams. The velocities have a median value of
$28\pm19$~~kms$^{-1}$ and are binned in groups of
$10$~~kms$^{-1}$.}\label{fig:vel}
\end{figure}

{The velocities of over one hundred jets are measured and a
histogram of their velocity distribution is shown in
Fig.~\ref{fig:vel}. The jets have a range of upward velocities,
$10-110$~km\,s$^{-1}$, with a median value of
$28\pm19$~~km\,s$^{-1}$.} The jets typically have velocities with
similar values reported previously for the type-I spicules, i.e.,
$10-40$~km\,s$^{-1}$ and also decelerate with height. The
deceleration of the jets is readily identified by the curved tracks
seen within Fig.~\ref{fig:xt_plots}. The deceleration is around
$0.03-0.23$~km\,s$^{-1}$, much less than the expected gravitational
deceleration. Similar behaviour is observed for spicules. However,
the jet plasma is rarely seen to return to the surface, fading from
the filter at the jet's peak height.

{On the right hand side of the sunspot, we see a similar picture.
However, more jets with velocities $>40$~~kms$^{-1}$ are observed
than compared to the left hand side. The high velocity jets are
observed typically when large-scale ejections of plasma occur, with
numerous jets driven simultaneously over a $6000$~km spatial scale.
This large-scale jet excitation only occurs on this side of the
sunspot and demonstrates that the driver of these events is
relatively localised (possibly connected with the overlying coronal
loops).}

{A number of receding jets are seen and down flows are also
observed. The number of these events that we can identify is
relatively small, namely 11 events in the 45 minute time series. The
events have an average velocity of $12\pm8$~km\,s$^{-1}$. This
velocity is comparable to down flows seen in TR spectral lines
(e.g., \citealp{MARetal2004}; \citealp{DAMetal2008};
\citealp{TERIetal2008}).}

\vspace{0.5cm}

\begin{figure}[!tbp]
\centering
\includegraphics[scale=0.5] 
{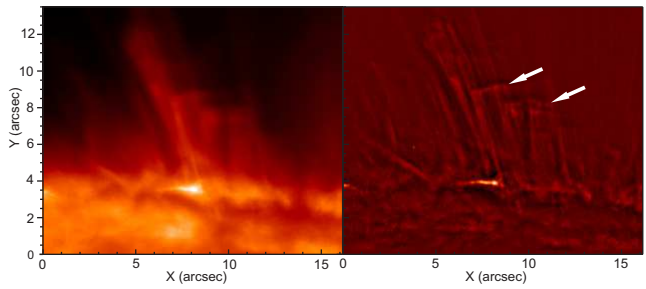}\caption{The \textit{left} column is a Ca II
image. The \textit{right} column shows an unsharp masked image of
the same region. The white arrows mark the enhanced emission at the
leading edges of the jet.}\label{fig:emis}
\end{figure}

One of the most interesting features is, unlike spicules, these jets
do not taper with height. It has been often reported that spicule
intensity decreases with height (e.g., \citealp{BEC1968}) and they
have a tapered appearance. In Figs.~\ref{fig:fov}, \ref{fig:pack}
and \ref{fig:emis} it can be observed that the jets have increased
emission along the leading edge of the groups, suggesting an
increase in temperature or density (or both). The increase in
emission at the leading edge is $\sim5\%$ of the emission of the the
trailing plasma. The enhancement is much clearer in the unsharp
masked image (Fig.~\ref{fig:emis}). The fact that only the leading
edge is enhanced suggests this is not a line of sight effect, i.e.,
a superposition jets. If this were the case, we would expect most or
all of the jet to show enhanced emission, not just the leading edge.

We suggest the cause of this enhanced emission is due to the leading
edge of the jet being a shock front. The shock would heat and
condense the plasma leading to enhanced emission. If sufficient
heating were to occur, then the ionization of Ca$^+$ ions would lead
to plasma disappearing from the SOT filter. This would explain why
some of the ejected plasma is not seen to return to the surface and
the lower emission of the returning plasma. However, the unshocked
plasma in front of the jet has to have a temperature below
$1-2\times10^5$~K. This is because the sound speed in the unshocked
plasma has to be less than the velocity of the jet
($30-40$~km\,s$^{-1}$) or else a shock cannot form. Models of
coronal loops suggest that relatively cool plasma, i.e. $T<0.5$~MK,
exists in the loop's legs even at coronal heights (e.g.,
\citealp{ASCetal2001}). The shock heating of chromospheric jets is
discussed in, e.g., \cite{HOL1982}; \cite{STEHOL1988};
\cite{STE2000}; \cite{JAMERD2002, JAMERD2003} \cite{MURetal2011}.

On viewing the TRACE $1550$~{\AA} and $1600$~{\AA} images
(Fig.~\ref{fig:fov}), the observed emission shows similar
geometrical features and can be seen to be co-spatial with the Ca II
H emission seen with SOT. The observed emission above the limb in
the TRACE bandpasses is most likely due to the contribution from C
IV, which is formed at $T\sim0.1$~MK. This would add weight to the
suggestion that shock fronts can from at the leading edge of the jet
and low level heating is occurring of the chromospheric jet plasma
to at least lower to mid TR temperatures.

{Unfortunately, the corresponding XRT, TRACE and SOHO/EIT data does
not provide sufficient temporal coverage to examine whether the jets
observed in Ca II H have counterparts at higher temperatures. Other
instruments, e.g. SUMER, CDS, also do not have data on this region.}
Such joint observations are undoubtedly needed to confirm whether
the observed jets have higher temperature counterparts and confirm
whether shock-waves (or other mechanisms) are heating the
chromospheric plasma.

\begin{figure}[!htbp]
\centering
\includegraphics[scale=0.7] 
{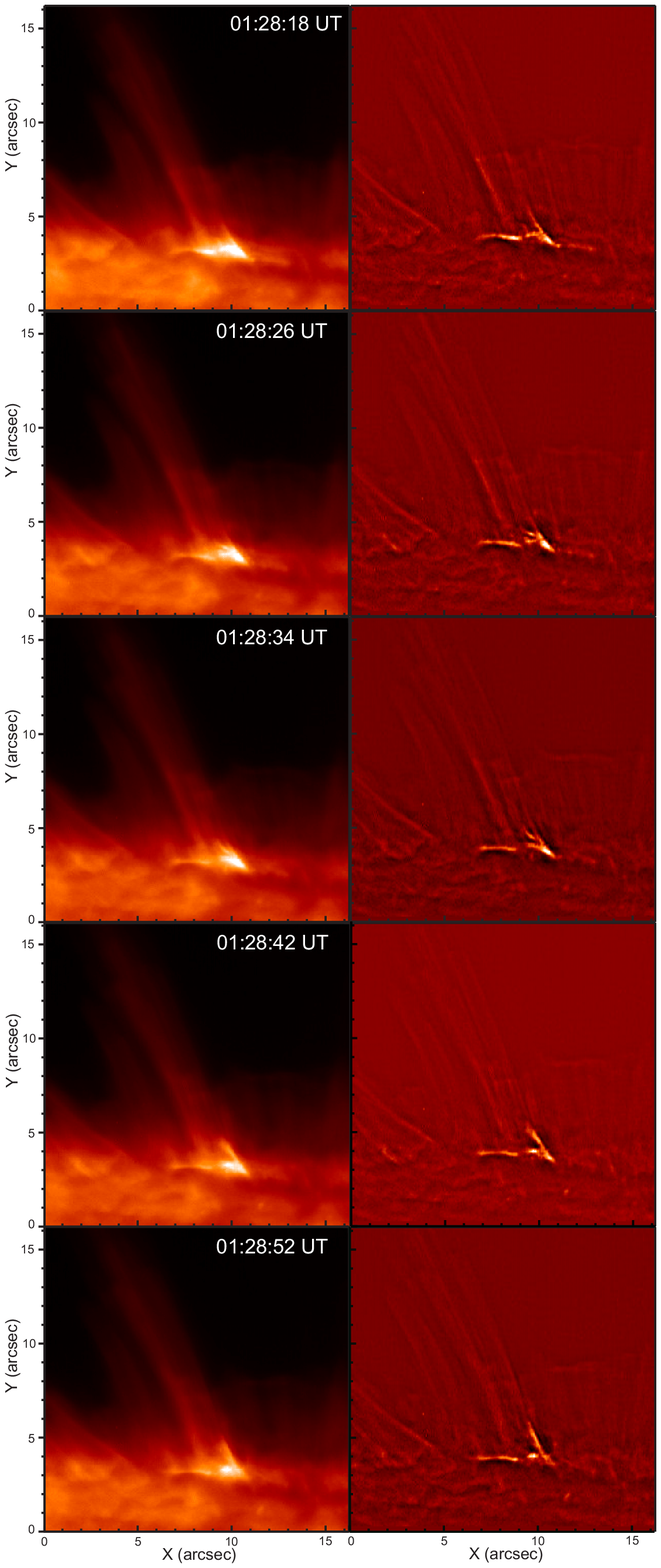}\caption{The \textit{left} column is Ca II images
of a sunspot showing plasma jets and the site of potential magnetic
reconnection. Each image is separated in time by $\sim8$~s. The
\textit{right} column shows the unsharp masked
images.}\label{fig:recon}
\end{figure}
\section{Excitation of the jets}
Now that we have described some of the observational features of the
jets, we demonstrate evidence that suggests potential excitation
mechanisms. In the quiet Sun it had been suggested that a mixture of
p-modes and granular buffeting are the main mechanisms responsible
for type-I spicule excitation (\citealp{DEPetal2004};
\citealp{ERDetal2007}), while reconnection is thought responsible
for driving type-II events (\citealp{DEPetal2007a}). We give
evidence that both methods are also at work in the formation of jet
events at the edge of sunspots.

\subsection{Reconnection in the sunspot}
Possible sites of magnetic reconnection can be identified in the
images. In particular from t=400~s to t=1100~s after the
observations begin, jet excitation appears to be associated with a
sustained bright region at the chromospheric level, $\sim3000$~km in
length. In Fig.~\ref{fig:recon} we display a series of images that
portray the typical behaviour in the bright region. In the first
image a short loop-like structure can be seen. In the next few
images, a portion of the loop is seen to rise upwards until it
protrudes above the low chromosphere and is aligned with existing
structures that support up-flowing plasma. This process is seen to
occur every $\sim50$~s while the bright region exists, suggesting
multiple reconnection events. The observed event is similar in
appearance to reconnecting chromospheric loops seen in large jet
events in the quiet chromosphere, e.g., \cite{LIUetal2009,
LIUetal2011}, \cite{MORetal2012}. The jets also display a transverse
motion across the sunspot, giving the appearance of the field lines
being dragged from right to left. This could be a sign of the
magnetic field reorganising above the reconnection site.

\subsection{Timescales of jet excitation}
Now, we investigate the typical timescales associated with the jets.
Using the time-distance diagrams shown in Fig.~\ref{fig:xt_plots},
we apply Fourier wavelet analysis at different heights. We find the
predominance of a 300~s timescale at the base of the cross-cuts,
both on the left and right hand sides of the sunspot. The presence
of timescales with periods of $400-500$~s are also significant but
have much less power than the $300$~s timescales. The power in the
$300$~s timescales drops off rapidly within the first $2$~Mm of the
cross-cuts. Higher in the atmosphere, the intensity is dominated by
$400-500$~s timescales. This timescale appears related to the
production of the jets that shoot material into the lower corona (
heights in excess of $6$~Mm). On Fig.~\ref{fig:xt_plots} we have
over plotted typical results from the wavelet. The wavelet scale
with the greatest power at a particular height is plotted. The shown
wavelet scale is plotted so that the zero-crossings correspond to
the height at which the wavelet is obtained.

We cannot rule out the inclusion of some plage region fibrils in
these cross-cuts contributing to the timescales. Fibrils are seen in
the data and a number occur near the sunspot. The fibrils are
expected to be a mixture of vertical and inclined structures,
however, previous observations (\citealp{DEPetal2007c}) suggest both
would not reach as high into the corona ($\sim1-4$~Mm) as the
sunspot jets. Hence, the fibrils should mainly be observed at the
foot points of the sunspot jets, i.e., at the bottom of the
cross-cuts. The inclusion of dynamic fibrils could be a reason why
$300$~s periods show significant power at the base of the
cross-cuts.

{The long timescales found here are similar to the periods of waves
reported in sunspots, e.g., \cite{THOetal1982}, \cite{OSHetal2001},
and in coronal loops in active regions, \cite{DEMetal2002}. It is
possible that waves could be exciting the jets either by non-linear
steepening to form shocks, which elevate the chromospheric material
(e.g., \citealp{HOL1982}, \citealp{DEPetal2004},
\citealp{ERDetal2007}) or exciting periodic reconnection (e.g.,
\citealp{HEGetal2009}, \citealp{MCLetal2010},
\citealp{MCLetal2012}).}

{We also note the $50$~s timescale associated with the discussed
reconnection event does not show significant power (i.e., $>99\%$
significance) in the wavelet time-series. It would appear the longer
time-scales (300-500~s) are associated with more powerful excitation
events, able to launch plasma to greater heights above the sunspot.
However, the observation of jet excitation on shorter timescales
shows that localised events with a range of timescales will also
play a significant role in determining the observed dynamics.}

\section{The chromosphere and TR above the sunspot}
Flows in sunspot and active region TRs have been studied in detail
using spectroscopy (e.g., \citealp{KJEetal1988},
\citealp{MARetal2004}, \citealp{WARetal2011}). In particular,
\cite{TIAetal2009} studied the profile of Hydrogen Lyman lines above
both plage and active regions. Typically, the Lyman-$\beta$ to
Lyman-$\epsilon$ show self-reversal over plage regions but not over
the sunspot umbra, suggesting a possible different pattern of flow
in the two regions. The differing magnetic structures in each region
probably plays a role in this phenomenon.

\cite{TIAetal2009} suggest that opacity over the sunspot is much
less than over the plage, leading to reduced absorption of the Lyman
emission on-disk. This in turn implied less chromospheric plasma
over sunspots. Further, they calculated electron densities over the
plage and umbra from a range of density sensitive lines pairs,
finding lower TR densities over the umbra when compared to the plage
region. It was suggested this was due to a {higher} and extend TR
over the sunspot in comparison to the plage TR.

\begin{figure}[!htbp]
\centering
\includegraphics[scale=0.8] 
{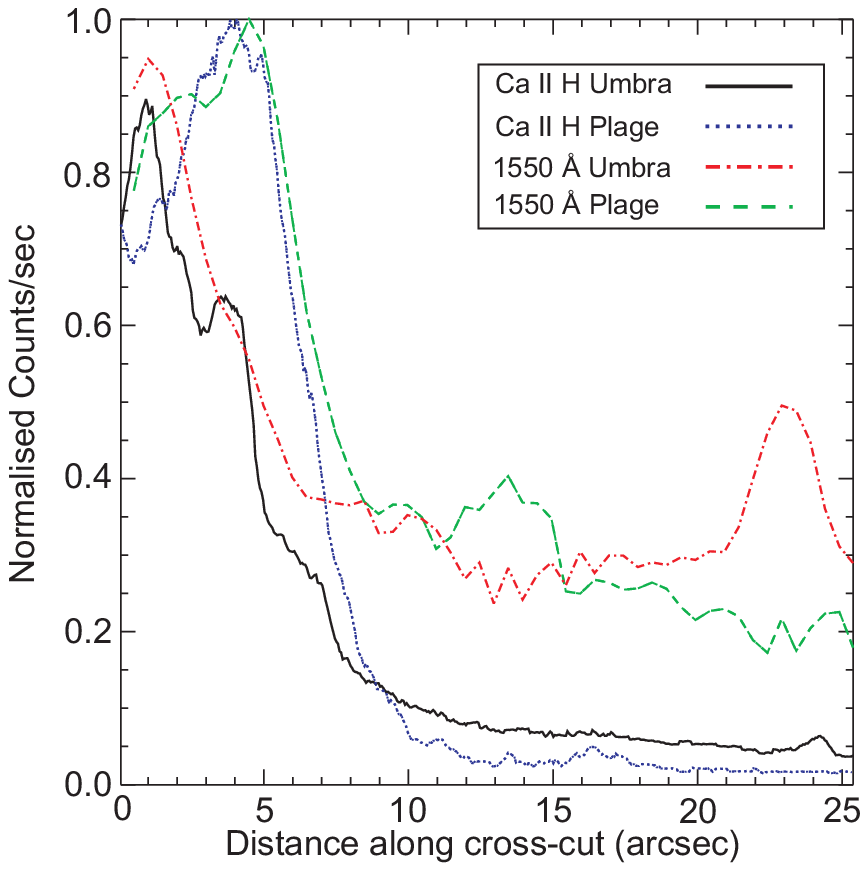}\caption{The flux profiles taken above plage and
umbral regions. The $1550$~{\AA} and Ca II H plage cross-cuts are
the dashed (green) and dotted (blue) lines, respectively. The
$1550$~{\AA} and Ca II H umbral cross-cuts are the dash-dot (red)
and solid (black) lines, respectively. The intensity have been
normalised to the largest counts/sec value in the umbral and plage
cross-cuts for each filter. The reduced emission in the Ca II H
umbral cross-cut at $\sim 3''$ along the slit is due to the Wilson
depression of the sunspot.}\label{fig:flux}
\end{figure}

{We can confirm this conjecture is indeed true. In the Hinode/SOT Ca
II H and the TRACE 1550~{\AA} images we have over plotted two solid
lines (Fig.~\ref{fig:fov}). The Ca II H cross-cuts are co-spatial
with the TRACE cross-cuts. The upper line marks the surrounding
plage region and the lower line marks the umbral region of the
sunspot. The flux along these cross-cuts is plotted in
Fig.~\ref{fig:flux} for the first image in the Ca II H time series
and the closest TRACE 1550~{\AA} image in time.}

{First, we can identify approximately where the solar surface is in
the flux profiles. The plage profiles in Fig.~\ref{fig:flux} show a
steep decrease in the flux after $\sim4$~'' along the cross-cut
signifying the reduction in photospheric contributions to the
observed emission. The umbral Ca II H flux also has a sharp decrease
at $1$~'' along the cross-cut before increasing again. This is due
to the Wilson depression which can be observed in
Fig.~\ref{fig:fov}. The 1550~\AA flux also begins to decrease at
$1$~'' implying the 1550~\AA emission is also sensitive to the
Wilson effect.}

{Above the solar surface, the TR emission is seen to be
significantly greater above the plage than the umbra to a distance
of $\sim9''$ along the cross-cut. This suggests the umbral TR is
less dense than the plage TR. After $\sim9''$, the emission is then
comparable until $\sim15''$ along the cross-cut when the plage TR
emission becomes less than the umbral TR. Hence, there is more TR
temperature material at greater heights above the umbra than the
plage. This demonstrates the higher and extended nature of the
umbral TR.}

{In Fig.~\ref{fig:flux} we also show the chromospheric Ca II H
emission along taken along the cross-cuts shown in
Fig.~\ref{fig:fov}. The chromospheric plage emission is greater than
the umbral emission until $9''$ along the cross-cut, suggesting a
denser chromosphere in the plage. However, the umbral chromosphere
can be seen to extend much higher than the plage region, with
greater umbral emission after $9''$ along the cross-cut. This should
be expected considering that the sunspot jets typically reach
greater heights than jets reported in the quiet Sun and plage
regions.}

The attributes of the umbral chromosphere and TR, i.e. less-dense,
higher and extended compared to the plage, can probably be put down
to the influence of the chromospheric jets. The sunspot jets appear
to be more dynamic than previously reported quiet Sun and plage
region jets. The jets observed here are able to distribute the
chromosphere much higher into the corona, also leaving the
chromospheric material thinner above the umbra. In turn this causes
the TR to be extended over greater heights above the umbra. This
will be the case whether the jets are responsible for generating the
TR, e.g., via shock heating, or simply just driving the TR material
to a greater height.

\section{Discussion and conclusion}

The jets observed around the edge of the sunspot display much
greater dynamism than jets observed previously in the quiet Sun and
plage regions. New jets appear almost continuously in groups,
ejecting large amounts of plasma along the field lines. Enhanced
emission at the leading edge of the jets and the lack of visible
down flowing plasma suggests that the jet material may be being
heated by shocks as it rises into the sunspot TR/corona. This claim
is supported by the presence of TR temperature material (C IV at
$T\sim0.1$~MK) in TRACE 1550~{\AA} and 1600~{\AA} filters that is
co-spatial with and shares similar geometric features to the Ca II H
emission. The shock wave-heating may not provide the hottest plasma
observed in active regions, i.e., temperatures in excess of $1$~MK,
but could provide a steady stream of warm plasma $T\sim0.1$~MK at
the footpoints of the active regions. {The heated jet plasma may add
to some of the observed transition region down flows above sunspots
(e.g., \citealp{KJEetal1988}, \citealp{BRYeta1999},
\citealp{TERIetal2008}, \citealp{TIAetal2009}) if the heated plasma
returns to the surface. However, the TR plasma could be heated
further, possibly to coronal temperatures, before it returns to the
surface (as suggested in \citealp{PNEKOP1978})}. Further studies
will have to focus on tracking the jet plasma as it heated to
determine its fate. This will require both high spatial and temporal
resolution observations covering a wide range of temperatures.

{We can obtain an estimate for the amount of chromospheric jet
plasma that needs to be heated to supply the TR down flows. The
technique used is similar to that in, e.g. \cite{ATHHOL1982},
\cite{DEPetal2009}.}

{The mass flux of chromospheric plasma is given by
$F_s=\epsilon\alpha_s n_{s} v_s$, where $\epsilon$ is the proportion
of mass heated to TR temperatures, $\alpha_s$ is the volume filled
by the jets, $n_s$ is the electron particle density and $v_s$ is the
up-flow speed of the chromospheric jets. The amount of chromospheric
material heated to TR temperatures has to at least equal the
downward mass flux seen in the TR lines, i.e.,
$F_{TR}=\alpha_{TR}n_{TR}v_{TR}$. The electron density in the TR
above sunspots is $\sim10^{10}$~cm$^{-3}$ (\citealp{TIAetal2009}),
with typical down flow velocities of 10-30~km\,s$^{-1}$
(\citealp{MARetal2004}; \citealp{DAMetal2008}), which gives a mass
flux of $F_{TR}=\alpha_{TR}2\pm1\times10^{16}$~cm$^{-2}$\,s$^{-1}$.}

{We have demonstrated tha the chromosphere above the sunspot umbra
is more extended and possibly less-dense than the plage
chromosphere. Because the emission measure $EM\propto n^2$, we can
provide a crude estimate for the umbral chromospheric electron
density. Assuming that temperatures and volumes are equal, the ratio
of emission in the plage to the umbra gives
$EM_{p}/EM_{U}=(n_p/n_u)^2$. For the Ca II emission we find
$EM_{p}/EM_{U}=2.1\pm0.5$ for heights between 1-2~Mm (1.3-3 '')
above the limb. The electron density of jets in plage regions is
$n_p\sim4\times10^{11}$~cm$^{-3}$ (\citealp{BEC1972}), hence
$n_u\sim2.5\times10^{11}$~cm$^{-3}$. Now $n_u=n_s$, hence the mass
flux associated with the chromospheric jets is
$F_s=\epsilon\alpha_s7.5\times10^{17}$~cm$^{-2}$\,s$^{-1}$, where
$v_s\sim30$~km\,s$^{-1}$. If we assume that the up-flowing/down
flowing plasma is confined to individual field lines, then the ratio
$\alpha_{TR}/\alpha_s$ will be a measure of the expansion of the
magnetic field with height. Estimated field strengths and changes in
magnetic field strength with height (\citealp{SOL2003}) give
$\alpha_{TR}/\alpha_s\sim2-3$. This gives $\epsilon\sim8\pm4\%$, so
the heating of only a fraction of the chromospheric jets will
replenish the mass lost from TR down flows.}

{Downflows are also observed in the Ca II H data. However, these are
the result of jet material returning to the surface under the
influence of gravity. No down flows are observed that cannot be
connected with the return of a jet.}

The data appears to demonstrate that a number of different
mechanisms are in action to drive the jets. Large ejections of
plasma, distinct magnetic topology and apparent re-organisation of
the magnetic field are observed suggesting magnetic reconnection
plays a role. The presence of timescales of $300$~s and $500$~s also
suggest that waves may play a role in excitation of the jets,
whether this is the formation of shocks to drive material upwards
(e.g., \citealp{HOL1982}; \citealp{DEPetal2004}) or driving periodic
reconnection (\citealp{HEGetal2009}; \citealp{MCLetal2010};
\citealp{MCLetal2012}).

The evidence for jet events similar to type-II spicules is limited
in these observations. We see a few fast moving events with speeds
$40-100$~km\,s$^{-1}$ which fall into the bracket of type-II. The
high-speed jets are only visible when a large ejection event occurs,
launching plasma up to $14$~Mm in to the sunspot corona.

{The reported velocities for up-flows in higher temperature lines
consistently exceeds $100$~km\,s$^{-1}$, e.g., \cite{TIAetal2011}.
This apparent discrepancy between our results might be able to be
explained in a number of ways. In the observations presented here,
the lack of evidence for events with rise-speeds in excess of
$100$~km\,s$^{-1}$ could be due to the observed transverse motions
obscuring any potential signals in time-distance plots. Certain jets
move transversally across the sunspot and pass through the
cross-cuts leaving an almost straight line in the time-distance
plots. We neglected any lines which gave velocities in excess of
$125$~km\,s$^{-1}$ as a precaution. The shear volume of jets at the
foot point also provides a significant restriction to observing
type-II like events due to the superposition of many jets. Very fast
type-II events, i.e., with velocities in excess of
$100$~km\,s$^{-1}$, may only exist in Ca II H for a short period of
time before significant plasma heating occurs. In such a scenario,
their signal will be hidden amongst the forest of jets observed near
the cross-cuts footpoints. Another option is that heated jets which
fade from Ca II H are accelerated and the rise velocity increases
with increasing temperature.}

In conclusion, we demonstrate that relatively slow moving jets
($v\sim30$~km\,s$^{-1}$) appear to experience shock heating. While
study of the fast moving ($v>40$~km\,s$^{-1}$) jet events has
demonstrated that the heating of plasma to coronal temperatures may
be associated with chromospheric phenomena (e.g.,
\citealp{DEPetal2011}), the role that {\lq slower\rq} jet events
could play in generating heated atmospheric plasma should not be
overlooked. It may be this role is restricted to low level heating
of the plasma up to transition region temperatures. However, even
such a low level process could {\lq reduce the load\rq} put upon
type-II events (or other heating mechanisms) in supplying heated
material to the transition region and corona of active regions.

\begin{acknowledgements}
The author thanks M. Ruderman, V. Fedun, G. Verth, A. Hillier, D.
Jess and R. Erd\'elyi for useful and insightful discussions. Thanks
goes to the Science and Technology Facilities Council (STFC), UK for
the financial support received. The analysis has been based on
observations done by the Hinode satellite. Hinode is a Japanese
mission developed and launched by ISAS/JAXA, with NAOJ as domestic
partner and NASA and STFC (UK) as international paertners. Is it
operated by these agencies in co-operation with ESA and NSC
(Norway).
\end{acknowledgements}
\bibliographystyle{aa}

\end{document}